# Enhanced Predictive Capability for Chaotic Dynamics by Modified Quantum Reservoir Computing

Longhan Wang, Yifan Sun[+] and Xiangdong Zhang[*]

Key Laboratory of advanced optoelectronic quantum architecture and measurements of Ministry of Education,

Beijing Key Laboratory of Nanophotonics & Ultrafine Optoelectronic Systems, School of Physics, Beijing

Institute of Technology, 100081, Beijing, China

*[+]Author to whom correspondence should be addressed. E-mail: zhangxd@bit.edu.cn; yfsun@bit.edu.cn

## Abstract

Deducing the states of spatiotemporally chaotic systems (SCSs) as they evolve in time is crucial for various applications. However, it is a dramatic challenge for generally achieving so due to the complexity of non-periodic dynamics and the hardness of obtaining robust solutions. In recent, there is a growing interest in approaching the problem using both classical and quantum machine learning methods. Although effective for predicting SCSs within a relative short time, the current schemes are not capable of providing robust solutions for longer time than training time. Here, we propose an approach for advancing the prediction of chaotic behavior. Our approach can be viewed as a novel quantum reservoir computing scheme, which can simultaneously capture the linear and the nonlinear features of input data and evolve under a modified Hamiltonian. Our work paves the way for a new avenue in handling SCSs.

## I. INTRODUCTION

The states of numerous natural systems evolve chaotically [1]. Obtaining the information of such systems is very much close to our daily life events, such as forecasting weathers, controlling unmanned aerial vehicles, etc. However, to characterize or predict the behavior of a chaotic system is generally hard. A robust solution to the systems is unable to obtained by using current techniques, leading to obstacles in both analytical and numerical treatments. A different idea for approaching the problem is establishing an approximator to the exact chaotic model by using limited data measured from the system. By far, significant efforts have been dedicated to the research direction, proposing the techniques like delay-coordinate embedding [2], which considers data from the current and previous time steps in a model. The development of machine learning [3,4] also prompt advance of such approximators deeply. Especially, the introduction of reservoir computing (RC) [5–33] significantly broadens the capability of dealing with chaotic systems, especially large SCSs [24,31,32]. Recently, Ref. [33] proposes the next-generation RC scheme based on nonlinear vector autoregression, achieving efficient data prediction with shorter training datasets.

With the advancement of quantum computing, researchers have also attempted to address SCS using quantum RC (QRC) [34–50]. In contrast to classical RC, QRC is expected to outperform its classical counterparts. a general problem for all machine learning methods is the requirements on training dataset. Despite the progress mentioned above, all these schemes are limited to predicting a small amount of data using a sufficiently large training dataset [24,31–34,49]. Thus, the cost for capturing the long-term dynamics of SCSs would increase quickly. To propose an approach for tackling the problem is tricky, because it is required to generate data as complicated as those obtained from chaotic systems.

In this work, we provide a new scheme of QRC to perform the SCS with higher predictive capability. In our approach, the quantum reservoir simultaneously captures both linear and nonlinear features of input data and then evolves under a modified Hamiltonian. As a result, not only do we have accomplished the prediction of data over an extended period using a short training dataset, but the prediction time also improves by more than one order compared with the state-of-the-art [33] under the same training time. This work opens up new windows for the widespread application of neural network-based quantum machine learning.

## II. SCHEME OF ENHANCED PREDICTIVE CAPABILITY FOR CHAOTIC FOR CHAOTIC DYNAMICS BY MODIFIED QRC

The traditional QRC scheme [34] comprises four main parts: the input signal, the quantum reservoir (QR), partial measurement or quantum state tomography of the QR, and the output signal. Although our scheme also consists of four main parts, a different strategy of constructing QR is employed. In our scheme, two pathways are used as illustrated in Fig. 1. One pathway (red arrows in Fig.1) still encodes the input signal to obtain a quantum state $\left| S_k \right\rangle$, in order to supply one part of the input information into the QR. The other pathway (blue arrows in Fig.1) displays the uniqueness of the scheme. Firstly, we define linear ($Q_{linear,k}$) and nonlinear ($Q_{nolinear,k}$) features at time step $k$ based on the d-dimensional input signal $\vec{x}_k = \left[x_1^{\ k}, x_2^{\ k} \cdots x_d^{\ k}\right]$. The $Q_{linear,k}$ is defined not only based on the current signal $\vec{x}_k$, but also taking into account previous input signals, and the $Q_{nolinear,k}$ is a nonlinear function of $Q_{linear,k}$. The defined details of $Q_{linear,k}$ and $Q_{nolinear,k}$ are provided in Appendix A.

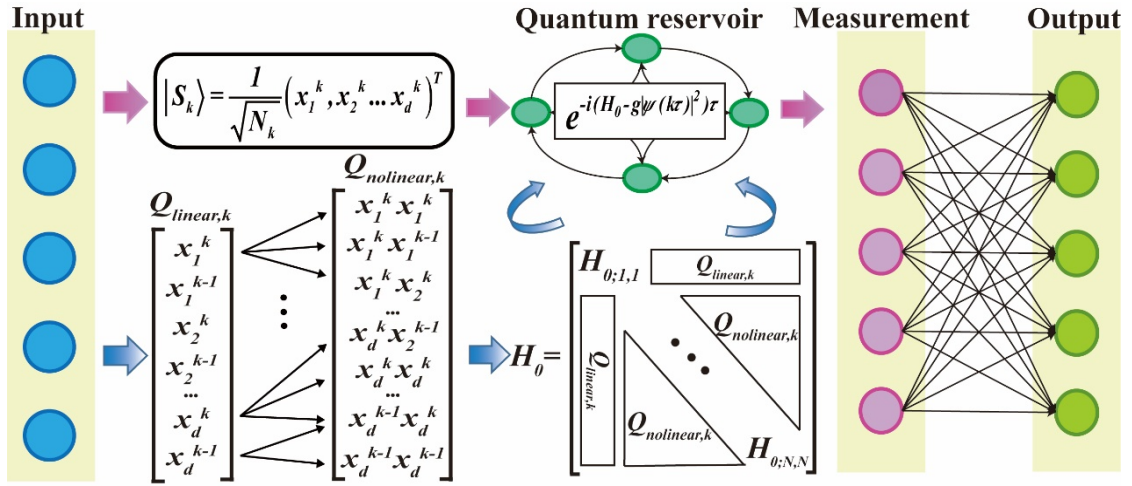

**Fig. 1.** Schematic of the QRCSPC. The $\left|S_k\right\rangle = \frac{1}{\sqrt{N_k}}\left[x_1^{\ k}, x_2^{\ k} \cdots x_d^{\ k}\right]^T$ represents the quantum state under amplitude encoding [51], where $N_k = x_1^{\ k2} + x_2^{\ k2} + \cdots + x_d^{\ k2}$ is the normalization coefficient. The $Q_{linear,k}$ and $Q_{nolinear,k}$ represent linear features and nonlinear features respectively, which are arranged sequentially onto the matrix elements of $H_0$. The output layer expresses the output signal as linear transformation of measurement signal in measurement layer.

The Hamiltonian $H_0$ can be given by $Q_{linear,k}$ and $Q_{nolinear,k}$. Specifically, the elements of $H_0$ above the main diagonal are given by the elements of $Q_{linear,k}$ and $Q_{nolinear,k}$. The elements $\left\{H_{0;i,i}\right\}_{i=1}^N$ on the main diagonal of $H_0$ are randomly set. Due to the Hermitian nature of $H_0$, the elements of $H_0$ below the main diagonal are obtained by transposing those above the diagonal, as shown in Fig. 1. Further details of $H_0$ are also provided in Appendix A. Then, the Hamiltonian corresponding to the QR in our scheme is given by

$$H = H_0 - g \left\| \psi\left(k\tau\right) \right\rangle \right|^2, \tag{1}$$

where $g \left\| \psi\left(k\tau\right) \right\rangle \right|^2$ represents a nonlinear self-potential term, $g$ is a constant, and $\left| \psi\left(k\tau\right) \right\rangle$ denotes a quantum state of QR under a input quantum state. In the case of a pure state, it corresponds to the squared modulus in vector form (diagonal elements in the density matrix). In the case of a mixed state, it corresponds to the diagonal elements in the density matrix as well. Thus, the unitary evolution in our QR can be described by $e^{-iH\tau}$.

Now, we discuss the process of this scheme for a sequence of $M$ input signal $\left\{ \vec{x}_k \right\}_{k=1}^{M}$. Similar to traditional scheme, our task is to utilize the quantum system to find a nonlinear function $\vec{y}_k = f\left( \left\{ \boldsymbol{x_1}^j, \boldsymbol{x_2}^j \cdots \boldsymbol{x_d}^j \right\}_{j=1}^{k} \right)$ such that the mean-square error between the practical output $\vec{y}_k$ and the target output $\overline{y}_k$ becomes minimum, here $\vec{y}_k = \left[ y_1^{\ k}, y_2^{\ k} \cdots y_l^{\ k} \right]$ and $\overline{y}_k = \left[ \overline{y}_1^{\ k}, \overline{y}_2^{\ k} \cdots \overline{y}_l^{\ k} \right]$. To achieve this goal, at each time $t = k\tau$, the input signal $\vec{x}_k$ is encoded into $\left| S_k \right\rangle$ through amplitude encoding [51], where data is encoded into the amplitudes of a quantum state. Denote its density matrix as $\rho_{in;k} = \left| S_k \right\rangle \left\langle S_k \right|$. Subsequently, $\rho_{in;k}$ is input into the QR, given by

$$\rho\left(k\tau\right) \rightarrow \rho_{in;k} \otimes tr_1\left( \rho\left(k\tau\right) \right), \tag{2}$$

where $tr_1\left( \rho\left(k\tau\right) \right)$ represents the partial trace of QR state $\rho\left(k\tau\right)$ with respect to $\rho_{in;k}$ and $tr_1\left( \rho\left(k\tau\right) \right)$ can be represented as

$$tr_1\left( \rho\left(k\tau\right) \right) = \sum_{m=0}^{d-1} \left[ \left\langle m \right| \otimes I\left( N/d \right) \right] \rho\left(k\tau\right) \left[ \left| m \right\rangle \otimes I\left( N/d \right) \right], \tag{3}$$

where $\left\langle m \right|$ represents the computational basis, $N$ denotes the dimension of the density matrix of the QR, and $I\left( N/d \right)$ represents the $\left( N/d \right)$-dimensional identity matrix. After inputting

$\rho_{in;k}$, the QR evolves under the Hamiltonian $H$ for a time interval $\tau$. Then, the density matrix of the QR can be given by

$$\rho\big((k+1)\tau\big) = e^{-iH\tau}\Big(\rho_{in;k} \otimes tr_1\big(\rho\big(k\tau\big)\big)\Big)e^{iH\tau}. \tag{4}$$

The unitary evolution of the QR is governed by the discrete nonlinear Schrödinger equation [52]. When $g > 0$, this nonlinear Schrödinger equation is equivalent to the well-known Gross−Pitaevskii equation [52]. It describes the interactions among bosons in a Bose-Einstein condensate in the mean-field limit. After obtaining the density matrix of the QR, we perform quantum state tomography [53] to estimate the complete state $\rho\big((k+1)\tau\big)$ of the QR. Suppose that one the measurement operators of quantum state tomography are defined as $\hat{\lambda}_o$. For a $N$-dimension density matrix $\rho\big((k+1)\tau\big)$, there are a total of $N^2$ measurement operators, denoted as $\big\{\hat{\lambda}_o\big\}_{o=0}^{N^2-1}$. The detailed description of $\hat{\lambda}_o$ is provided in Appendix B. The measurement signal is defined as the expectation value of the measurement operators, which is represented as $\big\langle\hat{\lambda}_o\big\rangle = tr\big(\hat{\lambda}_o\rho\big((k+1)\tau\big)\big)$. In this way, we ultimately obtain a vector $\vec{r}_k$ of measurement signal with size $N^2$. In general, performing complete tomography is an expensive task in terms of the number of measurements to reconstruct the density matrix. To address the reconstruction of high-dimensional density matrix, quantum compressed sensing techniques have been proposed [42]. These techniques are applicable to density matrices of arbitrary dimensions [54] and have been experimentally demonstrated for characterizing complex quantum systems [54,55].

Finally, we explain the training process of our scheme based on the measurement result $\vec{r}_k$. Here, $\vec{r}_k$ is sampled from the QR at each time step $k$. In this way, we obtain $M$ measurement signals denoted as $\big\{\vec{r}_k\big\}_{k=1}^M$. Then, $\big\{\vec{r}_k\big\}_{k=1}^M$ is represented as a data matrix and is denoted as $r = \big\{r_{kp}\big\}\big(1 \le k \le M, 1 \le p \le N^2\big)$. We also set $r_{k0} = 1$ as a bias term. Meanwhile, we use $y = \big\{y_{Lk}\big\}\big(1 \le L \le l, 1 \le k \le M\big)$ and $\bar{y} = \big\{\bar{y}_{Lk}\big\}\big(1 \le L \le l, 1 \le k \le M\big)$ to represent data matrices of the actual output signal and the target signal, both of which correspond to a matrix of size $l \times M$. In a least-squares sense, we match the actual output signal $y$ to the target signal $\overline{y}$. This problem corresponds to solving the following equation

$$\overline{y} = W_{out} r .\tag{5}$$

We assume that the length of the sequence $M$ is much larger than the dimension of the vector of measurement signal $\vec{r}_k$. That is, the above equation is overdetermined. Therefore, the weights that minimize the mean-square error is given by

$$W_{out} = \overline{y} r^T (r r^T)^{-1} .\tag{6}$$

Because the prediction task requires feedback from the system, during the training phase, we clamp the feedback from the system output and use the target outputs as the inputs, i.e., we set $\vec{x}_{k+1} = \overline{y}_k$. Based on the scheme described above, some complex tasks can be performed. In the following, we take some examples to verify the performance of our scheme.

## III. EXAMPLES

We consider two typical examples. One is a simplified weather system model [56] developed by Lorenz in 1963(Lorenz63). The other is the dynamics of a double-scroll electronic circuit [57]. Recently, both examples have been discussed using classical QR method in Ref. [33]. Now, we use the scheme to analyze these tasks and compare them with classical method.

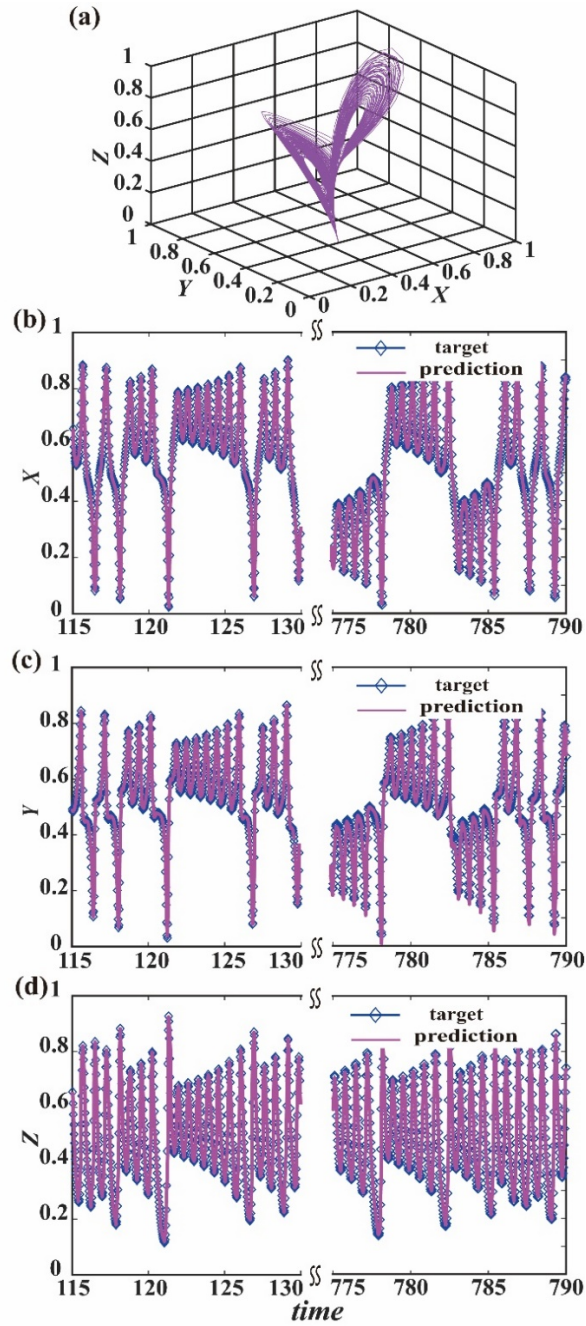

**Fig. 2.** (a) Schematic representation of dynamic System of standardized Lorenz63. The values of $X, Y, Z$ as a function of time about Lorenz63. (b), (c), (d). The values of $X, Y, Z$ of target (blue) and predication(purple) data during test phase.

For the Lorenz63, it consists of a set of three coupled nonlinear differential equations given by $\dot{X} = 10(Y - X)$, $\dot{Y} = X(28 - Z) - Y$ and $\dot{Z} = XY - 8Z / 3$ [56]. It displays deterministic chaos, sensitive dependence to initial conditions—the so-called butterfly effect—and the standardized phase space trajectory forms a strange attractor shown in Fig. 2(a). To perform the prediction task of Lorenz63 in Fig. 2(a), we generate data by numerically integrating with the time interval $\tau = 0.025$ like classical method. Based on the above sets, we obtain 31,600 standardized input data samples. The first 600 data samples are used for the washout, followed by 4,000 data samples for the training phase, and 27,000 data samples for the testing phase. At time $t = k\tau$, the state of Lorenz63 is represented as a column vector $\vec{x}_k = \left[ x^k, y^k, z^k \right]^T$. According to the column vector $\vec{x}_k$ and $\vec{x}_{k-1}$, the linear features $Q_{linear,k}$ are given with 6 components. Then, the nonlinear features $Q_{nolinear,k}$ are represented by $Q_{linear,k}$ and have 21 components. The detailed description can be found in Appendix C. In addition, we also set a constant $C = 1$. In this way, we can obtain a total of 28 components. Next, we arrange them in the upper part of the main diagonal of the Hamiltonian matrix $H_0$ of size $8 \times 8$. More detailed descriptions can be also found in Appendix C. To ensure that $\left( N / d \right)$ is an integer, we encode $\vec{x}_k$ as a quantum state $\left| S_k \right\rangle = \frac{1}{\sqrt{N_k}} \left[ x^k, y^k, z^k, 0 \right]^T$ and input it into the QR.

According to Eqs. (1)-(8), we perform the forecasting task by numerical simulation. The results for variables X, Y and Z are shown in Fig. 2(b), 2(c) and 2(d), respectively, which plot the target datasets (blue) and prediction datasets (purple) during the testing phase. Due to the large amount of data shown in the above figure, we only show the data from $t = 115$ to $t = 130$ and the data from $t = 775$ to $t = 790$. The other data are provided in Appendix D. It is seen clearly that the agreement between prediction datasets and target datasets is very well. We computed both the normalized root-mean-square error (NRMSE) [33], $\sqrt{\dfrac{\sum_{k=1}^{M} \left( \bar{y}_l^k - y_l^k \right)^2}{\sum_{k=1}^{M} \bar{y}_l^{k2}}}$, largest Lyapunov exponents [58] and power spectrum density (PSD) [58,59]. NRMSE are only $3.7 \times 10^{-3}$, $14.6 \times 10^{-3}$ and $7.5 \times 10^{-3}$ for three cases, respectively. The largest Lyapunov exponents in the target system and the prediction system, are 0.8693, 0.8543 respectively. The method of measuring the largest Lyapunov exponent is shown in Appendix G. PSD are provided in Appendix E. In Appendix E, we compared two different PSD methods. In the first method, PSD is defined as $20 \log_{10} \left( 2 \left| \text{FFT} \left( \vec{y}_k \right) \right| \right)$ where $\text{FFT} \left( \vec{y}_k \right)$ is the complex Fourier spectrum of the state evolution [58]. Numerical results in Fig. E1 (a) show that the PSD values of the target system and the prediction system are very close, thereby indirectly indicating consistency between the target data and the predicted data. In the second method, PSD is defined as $\left| FFT(Z(k)) \right|^2$, where $Z(k)$ denotes the values of Z variables at time step $k$ [59]. Similarly, numerical results in Fig. E2 (a) show that the PSD values of the target system and the prediction system are very close. This leads to the same conclusion as the first method that the target data and the predicted data are consistent. For comparison with the classical scheme [33], we can use the data sampled from 100 seconds of chaotic dynamics to predict the dynamics data in the following 675 seconds, which is 13.5 times higher than the corresponding classical results within the same training time.

For the double-scroll electronic circuit, it also consists of a set of three coupled nonlinear differential equations given by $\dot{V_1} = V_1/D_1 - \Delta V/D_2 - 2D_5 \sinh(D_4 \Delta V)$, $\dot{V_2} = \Delta V/D_2 + 2D_5 \sinh(D_4 \Delta V) - I$ and $\dot{I} = V_2 - D_3 I$ [57]. Here, $D_1 = 1.2$, $D_2 = 3.44$, $D_3 = 0.193$, $D_4 = 11.6$, $D_5 = 2.25 \times 10^{-5}$ and $\Delta V = V_1 - V_2$. It demonstrates the chaotic behavior of an electrical circuit system as shown in Fig. 3(a). To perform the prediction task of electrical circuit system in Fig. 3a, we generate data by numerically integrating with the time interval $\tau = 0.25$, which is also identical with that in Ref. [33]. Based on the above sets, we obtain 65,600 standardized input data samples. The first 600 data samples are also used for the washout, followed by 5,000 data samples for the training phase, and 60,000 data samples for the testing phase. At time $t = k\tau$, the state of the double-scroll electronic circuit is represented as a column vector $\vec{x}_k = \left[ V_1^k, V_2^k, I^k \right]^T$. According to the column vector $\vec{x}_k$ and $\vec{x}_{k-1}$, the $Q_{linear,k}$ are also given with 6 components. In such a case, the $Q_{nolinear,k}$ has 56 components, the detailed information is also given in Appendix C. That is, a total of 62 components are arranged in the upper part of the main diagonal of the Hamiltonian matrix $H_0$ of size $16 \times 16$. Similar to the above example, we also perform the forecasting task by numerical simulation. The results are shown in Fig. 3(b), 3(c) and 3(d). In Fig.3, the data from $t = 1400$ to $t = 1600$ and the data from $t = 16200$ to $t = 16400$ are only shown, the other data are also provided in Appendix D. We also computed both NRMSE, largest Lyapunov exponents and PSD. NRMSE are only $8.9 \times 10^{-3}$, $11.7 \times 10^{-3}$ and $8.8 \times 10^{-3}$ for three cases, respectively. The largest Lyapunov exponents are $0.0948$, $0.0859$ respectively. PSD are also provided in Appendix E. In such a case, the prediction time and training time are 15,000 seconds and 1,250 seconds, respectively. The predicted time is 24 times higher than the classical case within the same training time, which exhibit quantum advantage very well.

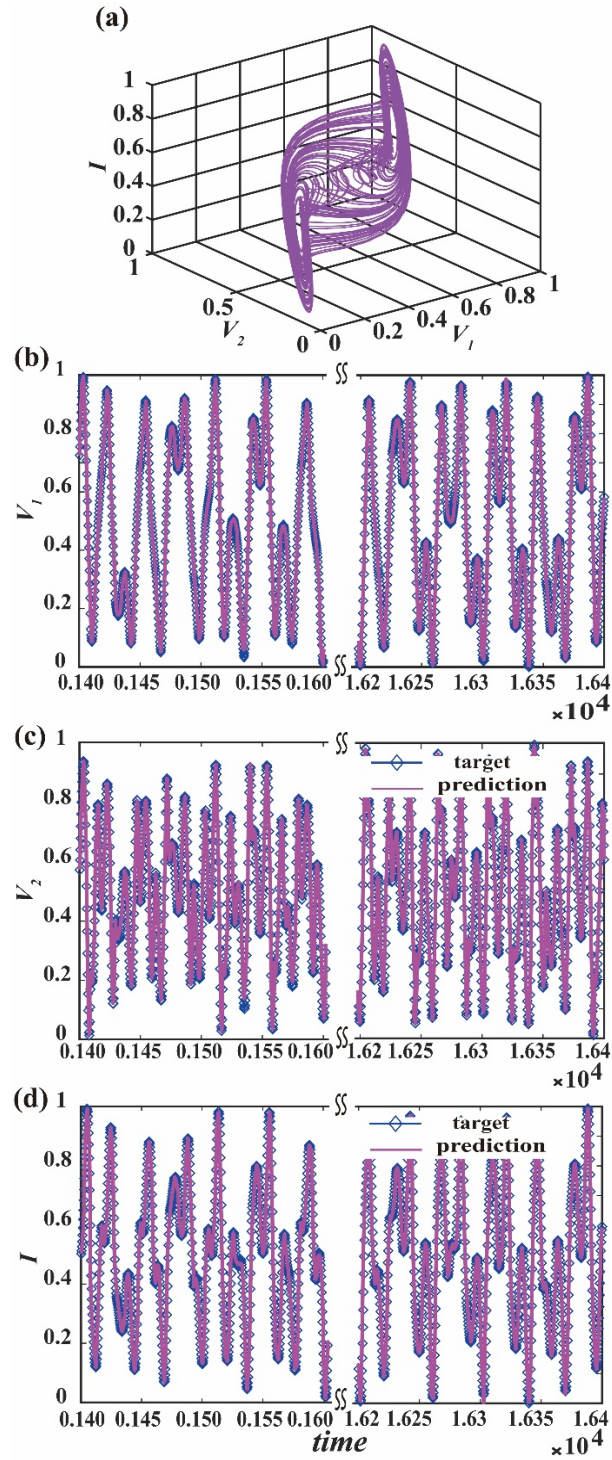

**Fig. 3.** (a) Schematic representation of standardized double-scroll electronic circuit of dynamic System. The values of $V_1, V_2$, and $I$ as a function of time about double-scroll electronic circuit. (b), (c), (d). The values of $V_1, V_2$, and $I$ of target (blue) and predication(purple) data during test phase.

## IV. DISCUSSION AND CONCLUSION

In the above study, we have only provided two examples. In fact, our scheme can be used to study a wide range of tasks. For some simple tasks, which the previous QRC scheme can work, this scheme still works. For more complex multi-attractor chaotic systems, even the rapid scrambling nature of black holes [60,61], their dynamic behavior can also be predicted by using our scheme, if linear features and the order of nonlinear features are modified according to the symmetry of the attractors and the size of time intervals.

In addition, we would like to emphasize that the theoretical framework we have proposed is, in principle, implementable in some real physical systems [52,62–64]. For example, if we construct evanescently coupled waveguide systems with Kerr nonlinearity as described in Ref. [52], and they are arranged in a two-dimensional array, the corresponding Hamiltonian with Eq. (1) can be obtained. Furthermore, our approach is expected to be implemented experimentally. More details about the experimental protocol are shown in APPENDIX F. This means that our work can indeed provide new avenues for performing complex machine learning tasks with physical devices.

In our future research, we plan to dig into the foundations underlying the capability of the modified QRC, by using the tools such as information processing capacity (IPC) [65]. IPC is an important tool for analyzing the memory character of a dynamical process, which has been applied to benchmarking quantum reservoirs in recent [66-71]. In the previous studies, the IPC is calculated under the independent and identically distributed input. How to properly set the nonlinear feature in such a case for obtaining an effective IPC would be a problem-to-solve for our proposal.

## ACKNOWLEDGMENTS

This work is supported by the National Key R & D Program of China under Grant No. 2022YFA1404900 and the National Natural Science Foundation of China (12234004).

## APPENDIX A: THE LINEAR FEATURES $Q_{linear,k}$ AND THE NONLINEAR FEATURES $Q_{nolinear,k}$ AND HAMILTONIAN $H_0$

We directly define the linear features $Q_{linear,k}$ from the d-dimensional input signal $\vec{x}_k = \left[ x_1^{\ k}, x_2^{\ k} \cdots x_d^{\ k} \right]$. It is composed of the input signal $\vec{x}_k$ at the current time step and the $(u-1)$ previous time steps, where the time gap between steps is $(q-1)$ [33]. This can be expressed by

$$Q_{linear,k} = \vec{x}_k \oplus \vec{x}_{k-q} \oplus \vec{x}_{k-2q} \oplus \cdots \oplus \vec{x}_{k-(u-1)q}, \tag{A1}$$

where $\oplus$ denotes the vector concatenation operation. Additionally, the nonlinear features $Q_{nolinear,k}$ are composed of the linear features $Q_{linear,k}$. An $sth$-order polynomial nonlinear features can be expressed by

$$Q_{nolinear,k}^s = Q_{linear,k} \boxtimes Q_{linear,k} \boxtimes \cdots \boxtimes Q_{linear,k}, \tag{A2}$$

where $Q_{linear,k}$ is repeated $S$ times, and $Q_{linear,k} \boxtimes Q_{linear,k}$ represents the vector of size $(du)(du+1)/2$ by taking all the elements of the outer product between $Q_{linear,k}$, removing duplicate elements. Logically, we can obtain $sth$-order polynomial nonlinear features $Q_{nolinear,k}^s$. We then arrange linear features $Q_{linear,k}$ and the nonlinear features $Q_{nolinear,k}$ from left to right, top to bottom in the upper part of the main diagonal of the Hamiltonian $H_0$. the elements $\left\{ H_{0;i,i} \right\}_{i=1}^{N}$ on the main diagonal of are randomly set. Due to the Hermitian property of $H_0$, we represent it as a symmetric matrix in Fig.A1. Assuming that the sum of the vector elements in $Q_{linear,k}$ and $Q_{nolinear,k}$ is $Q_k$, the relationship between the dimension $N$ of Hamiltonian $H_0$ and $Q_k$ should satisfy $(N-1)(N-2)/2 < Q_k \leq N(N-1)/2$. When $Q_k < N(N-1)/2$, the remaining elements are filled with constant.

**Fig.A1.** Schematic of the $H_0$. The $Q_{linear,k}$ and $Q_{nolinear,k}$ are arranged sequentially onto the matrix elements of $H_0$.

## APPENDIX B: QUANTUM STATE TOMOGRAPHY

Quantum state tomography (please see Ref. [53] in the main text for more details) is the experimental method of estimating a density matrix. Next, we will provide a detailed explanation of how to perform quantum state tomography for a quantum state with $d$ levels. First, we prepare the generators of the $SU(d)$ group for the $d$-dimensional Hilbert space, which allows us to construct the density matrix corresponding to the $d$-level quantum state. The generators of the $SU(d)$ group can be conveniently constructed using $d$-dimensional elementary matrices $\left\{ E_l^j \mid l, j = 1, \cdots, d \right\}$. Elementary matrices [51] can be given by

$$\left( E_l^j \right)_{\mu\nu} = \delta_{\nu l} \delta_{\mu j}, 1 \le \mu, \nu \le d \ , \tag{B1}$$

where $\delta$ represents the Dirac $\delta$ function. In the elementary matrices mentioned above, only the element in the $lth$ row and $jth$ column is 1, while all other elements are 0. From these elementary matrices, we can obtain $d(d-1)$ traceless matrices, which can be given by

$$\Lambda_l^j = E_l^j + E_j^l, \tag{B2}$$

$$\chi_l^j = -i \left( E_l^j - E_j^l \right). \tag{B3}$$

These are the off-diagonal generators of the $SU(d)$ group. Additionally, we give $(d-1)$ traceless but diagonal generators, which can be given by

$$\gamma_n^n = \sqrt{\frac{2}{n(n+1)}} \left[ \sum_{l=1}^{n} E_l^l - nE_{l+1}^{l+1} \right]. \tag{B4}$$

This way, we can obtain a total of $d^2 - 1$ generators. Next, we define the matrix forms corresponding to the measurement operators

$$\lambda_{(l-1)^2 + 2(j-1)} = \Lambda_l^j, \tag{B5}$$

$$\lambda_{(l-1)^2 + 2j-1} = \chi_l^j, \tag{B6}$$

$$\lambda_{l^2-1} = \gamma_{l-1}^{l-1}. \tag{B7}$$

The $d$-dimensional matrices in Eq. (B5-B7) and an $d$-dimensional identity matrix form a complete set of measurement operators. A density matrix of $d$-dimension can be expressed as a linear combination of these measurement operators, represented by

$$\rho_d = \frac{1}{d} \sum_{m=0}^{d^2-1} \alpha_m \hat{\lambda}_m d^2 - 1. \tag{B8}$$

To ensure the normalization of the diagonal elements of the density matrix $\rho_d$, we require

$\alpha_0 = 1$. Furthermore, to ensure that $tr\left( \rho_d^2 \right) \leq 1$, we require $\sum_{m=1}^{d^2-1} \alpha_m^2 \leq d(d-1)/2$.

**APPENDIX C: HAMILTONIAN $H_0$ IN NUMERICAL SIMULATION**

We have set the Hamiltonian $H_0$ for both the Lorenz63 and double-scroll electronic circuit tasks. For the Lorenz63, based on the linear and nonlinear features as shown in Eq. (A1) (A2), we can give the following equations

$$Q_{linear,k} = \left[ x_k, y_k, z_k, x_{k-1}, y_{k-1}, z_{k-1} \right], \tag{C1}$$

$$Q_{nonlinear,k} = \begin{bmatrix} x_k x_k, x_k y_k, x_k z_k, x_k x_{k-1}, x_k y_{k-1}, x_k z_{k-1}, y_k y_k, y_k z_k, y_k x_{k-1}, y_k y_{k-1}, y_k z_{k-1}, z_k z_k, \\ z_k x_{k-1}, z_k y_{k-1}, z_k z_{k-1}, x_{k-1} x_{k-1}, x_{k-1} y_{k-1}, x_{k-1} z_{k-1}, y_{k-1} y_{k-1}, y_{k-1} z_{k-1}, z_{k-1} z_{k-1} \end{bmatrix}. \tag{C2}$$

Additionally, we have also set a constant $C = 1$. With this, we arrange these 28 components from left to right, top to bottom in the upper part of the main diagonal of the Hamiltonian $H_0$ of size $8 \times 8$. The elements $\left\{ H_{0;i,i} \right\}_{i=1}^{8}$ on the main diagonal of $H_0$ are set from top-left to bottom-right as $200, 400, 600, 800, 800, 600, 400, 200$. Due to the Hermitian property of $H_0$, we represent it as a symmetric matrix. We then give the $H_0$ by

$$H_{0;8,8} = \begin{bmatrix} 200 & C & x_k & y_k & z_k & x_{k-1} & y_{k-1} & z_{k-1} \\ C & 400 & x_k x_k & x_k y_k & x_k z_k & x_k x_{k-1} & x_k y_{k-1} & x_k z_{k-1} \\ x_k & x_k x_k & 600 & y_k y_k & y_k z_k & y_k x_{k-1} & y_k y_{k-1} & y_k z_{k-1} \\ y_k & x_k y_k & y_k y_k & 800 & z_k z_k & z_k x_{k-1} & z_k y_{k-1} & z_k z_{k-1} \\ z_k & x_k z_k & y_k z_k & z_k z_k & 800 & x_{k-1} x_{k-1} & x_{k-1} y_{k-1} & x_{k-1} z_{k-1} \\ x_{k-1} & x_k x_{k-1} & y_k x_{k-1} & z_k x_{k-1} & x_{k-1} x_{k-1} & 600 & y_{k-1} y_{k-1} & y_{k-1} z_{k-1} \\ y_{k-1} & x_k y_{k-1} & y_k y_{k-1} & z_k y_{k-1} & x_{k-1} y_{k-1} & y_{k-1} y_{k-1} & 400 & z_{k-1} z_{k-1} \\ z_{k-1} & x_k z_{k-1} & y_k z_{k-1} & z_k z_{k-1} & x_{k-1} z_{k-1} & y_{k-1} z_{k-1} & z_{k-1} z_{k-1} & 200 \end{bmatrix}. \quad \text{(C3)}$$

For the double-scroll electronic circuit, based on the linear and nonlinear features as shown in Eq. (A1) and (A2), we can give the following equations

$$Q_{linear,k} = \begin{bmatrix} V_{1,k}, V_{2,k}, I_k, V_{1,k-1}, V_{2,k-1}, I_{k-1} \end{bmatrix}, \quad \text{(C4)}$$

$Q_{nolinear,k}$

$$= \begin{bmatrix} V_{1,k}V_{1,k}V_{1,k}, V_{1,k}V_{1,k}V_{2,k}, V_{1,k}V_{1,k}I_k, V_{1,k}V_{1,k}V_{1,k-1}, V_{1,k}V_{1,k}V_{2,k-1}, V_{1,k}V_{1,k}I_{k-1}, V_{1,k}V_{2,k}V_{2,k}, V_{1,k}V_{2,k}I_k, V_{1,k}V_{2,k}V_{1,k-1}, \\ V_{1,k}V_{2,k}V_{2,k-1}, V_{1,k}V_{2,k}I_{k-1}, V_{1,k}I_kI_k, V_{1,k}I_kV_{1,k-1}, V_{1,k}I_kV_{2,k-1}, V_{1,k}I_kI_{k-1}, V_{1,k}V_{1,k-1}V_{1,k-1}, V_{1,k}V_{1,k-1}V_{2,k-1}, V_{1,k}V_{1,k-1}I_{k-1}, \\ V_{1,k}V_{2,k-1}V_{2,k-1}, V_{1,k}V_{2,k-1}I_{k-1}, V_{1,k}I_{k-1}I_{k-1}, V_{2,k}V_{2,k}V_{2,k}, V_{2,k}V_{2,k}I_k, V_{2,k}V_{2,k}V_{1,k-1}, V_{2,k}V_{2,k}V_{2,k-1}, V_{2,k}V_{2,k}I_{k-1}, V_{2,k}I_kI_k, \\ V_{2,k}I_kV_{1,k-1}, V_{2,k}I_kV_{2,k-1}, V_{2,k}I_kI_{k-1}, V_{2,k}V_{1,k-1}V_{1,k-1}, V_{2,k}V_{1,k-1}V_{2,k-1}, V_{2,k}V_{1,k-1}I_{k-1}, V_{2,k}V_{2,k-1}V_{2,k-1}, V_{2,k}V_{2,k-1}I_{k-1}, \\ V_{2,k}I_{k-1}I_{k-1}, I_kI_kI_k, I_kI_kV_{1,k-1}, I_kI_kV_{2,k-1}, I_kI_kI_{k-1}, I_kV_{1,k-1}V_{1,k-1}, I_kV_{1,k-1}V_{2,k-1}, I_kV_{1,k-1}I_{k-1}, I_kV_{2,k-1}V_{2,k-1}, I_kV_{2,k-1}I_{k-1}, \\ I_kI_{k-1}I_{k-1}, V_{1,k-1}V_{1,k-1}V_{1,k-1}, V_{1,k-1}V_{1,k-1}V_{2,k-1}, V_{1,k-1}V_{1,k-1}I_{k-1}, V_{1,k-1}V_{2,k-1}V_{2,k-1}, V_{1,k-1}V_{2,k-1}I_{k-1}, V_{1,k-1}I_{k-1}I_{k-1}, V_{2,k-1}V_{2,k-1}V_{2,k-1}, \\ V_{2,k-1}V_{2,k-1}I_{k-1}, V_{2,k-1}I_{k-1}I_{k-1}, I_{k-1}I_{k-1}I_{k-1} \end{bmatrix}. $$

(C5)

With this, we arrange these 62 components from left to right, top to bottom in the upper part of the main diagonal of the Hamiltonian, whose matrix size is $16 \times 16$. We have only used the Hamiltonian sized by $14 \times 14$, and the remaining elements are filled with zeros. Since the 62 components are fewer than the 66 elements in the upper part of the main diagonal of $H_0$, we fill the remaining parts with the constant 10. The elements $\{H_{0;i,i}\}_{i=1}^{14}$ on the main diagonal of are set from top-left to bottom-right as $4000$. Due to the Hermitian property of $H_0$, we represent it as a symmetric matrix. We then give the $H_0$ by

$H_{0;16,16}$

$$
\begin{bmatrix}
0 & 0 & 0 & 0 & 0 & 0 & 0 & 0 & 0 & 0 & 0 & 0 & 0 & 0 & 0 & 0 & 0 \\
0 & 0 & 0 & 0 & 0 & 0 & 0 & 0 & 0 & 0 & 0 & 0 & 0 & 0 & 0 & 0 & 0 \\
0 & 0 & 4000 & V_{1,k} & V_{2,k} & I_k & V_{1,k-i} & V_{2,k-i} & I_{k-i} & V_{1,k}V_{1,k}V_{2,k} & V_{1,k}V_{2,k}I_k & V_{2,k}V_{2,k}V_{2,k} & V_{1,k}V_{2,k}V_{2,k-i} & 0 & 0 \\
0 & 0 & V_{1,k} & 4000 & V_{1,k}V_{2,k}I_{k-i} & V_{1,k}V_{2,k}V_{2,k} & V_{1,k}V_{2,k}I_k & V_{1,k}V_{2,k}V_{2,k-i} & V_{1,k}V_{2,k}V_{2,k-i} & V_{1,k}V_{2,k}I_{k-i} & V_{1,k}I_kI_k & V_{1,k}I_kV_{1,k-i} & V_{1,k}I_kV_{2,k-i} & V_{1,k}I_kI_{k-i} & 0 & 0 \\
0 & 0 & V_{2,k} & V_{1,k}V_{2,k}I_{k-i} & 4000 & V_{1,k}V_{2,k}V_{2,k} & V_{1,k}V_{2,k}V_{2,k-i} & V_{1,k}V_{2,k}V_{2,k-i} & V_{1,k}V_{2,k-i}I_k & V_{1,k}V_{2,k}I_{k-i} & V_{2,k}I_kI_k & V_{2,k}I_kV_{1,k-i} & V_{2,k}I_kV_{2,k-i} & V_{2,k}I_kI_{k-i} & 0 & 0 \\
0 & 0 & I_k & V_{1,k}V_{2,k}V_{2,k} & V_{1,k}V_{2,k}V_{2,k} & 4000 & V_{2,k}V_{2,k}V_{2,k-i} & V_{2,k}V_{2,k}I_{k-i} & V_{2,k}I_kI_k & V_{2,k}I_kI_k & V_{2,k}I_kI_k & V_{2,k}I_kV_{1,k-i} & V_{2,k}I_kV_{2,k-i} & V_{2,k}I_kI_{k-i} & 0 & 0 \\
0 & 0 & V_{1,k-i} & V_{1,k}V_{2,k}I_k & V_{1,k}V_{2,k}V_{2,k} & V_{2,k}V_{2,k}I_{k-i} & 4000 & I_kI_kI_{k-i} & I_kV_{1,k-i}V_{1,k-i} & I_kV_{1,k-i}V_{2,k-i} & I_kV_{1,k-i}I_{k-i} & I_kV_{1,k-i}I_{k-i} & I_kV_{1,k-i}V_{2,k-i} & I_kV_{2,k-i}I_{k-i} & 0 & 0 \\
0 & 0 & V_{2,k-i} & V_{1,k}V_{2,k}V_{2,k-i} & V_{1,k}V_{2,k}V_{2,k-i} & V_{2,k}V_{2,k}I_{k-i} & 4000 & I_kI_kI_{k-i} & V_{2,k}V_{2,k-i}I_{k-i} & I_kI_kI_{k-i} & I_kI_kI_{k-i} & V_{1,k-i}I_{k-i}I_{k-i} & I_kI_{k-i}I_{k-i} & I_kI_kI_{k-i} & 0 & 0 \\
0 & 0 & I_{k-i} & V_{1,k}V_{2,k}I_{k-i} & V_{1,k}V_{2,k}V_{2,k-i} & V_{2,k}I_kI_k & V_{2,k}V_{2,k-i}I_k & V_{2,k}V_{2,k-i}I_{k-i} & 4000 & V_{1,k-i}V_{2,k-i}I_{k-i} & V_{1,k-i}I_{k-i}I_{k-i} & V_{2,k-i}V_{2,k-i}I_{k-i} & V_{2,k-i}V_{2,k-i}I_{k-i} & V_{2,k-i}V_{2,k-i}I_{k-i} & 0 & 0 \\
0 & 0 & V_{1,k}V_{1,k}V_{2,k} & V_{1,k}V_{2,k}I_k & V_{1,k}V_{2,k}I_{k-i} & V_{2,k}I_kI_k & V_{2,k}I_kI_k & I_kI_kI_k & I_kV_{1,k-i}V_{2,k-i} & 4000 & V_{1,k-i}V_{2,k-i}I_{k-i} & V_{1,k-i}I_{k-i}I_k & 10 & 10 & 0 & 0 \\
0 & 0 & V_{1,k}V_{2,k}I_k & V_{1,k}I_kI_k & V_{2,k}V_{2,k}V_{2,k} & V_{2,k}I_kI_k & I_kI_kI_k & I_kV_{1,k-i}I_{k-i} & I_kV_{1,k-i}I_k & V_{1,k-i}I_{k-i}I_k & 4000 & 10 & 10 & 0 & 0 \\
0 & 0 & V_{2,k}V_{2,k}V_{2,k} & V_{1,k}I_kV_{2,k-i} & V_{2,k}V_{2,k}V_{2,k} & V_{2,k}I_kV_{2,k-i} & I_kI_kI_{k-i} & I_kV_{2,k-i}I_{k-i} & V_{1,k-i}V_{2,k-i}I_{k-i} & 10 & 10 & 10 & 4000 & 0 & 0 & \cdot \\
0 & 0 & 0 & 0 & 0 & 0 & 0 & 0 & 0 & 0 & 0 & 0 & 0 & 0 & 0 \\
0 & 0 & 0 & 0 & 0 & 0 & 0 & 0 & 0 & 0 & 0 & 0 & 0 & 0 & 0
\end{bmatrix}
$$

(C6)

## APPENDIX D: THE RESULTS OF NUMERICAL SIMULATION FOR THE LORENZ63 AND THE DOUBLE-SCROLL ELETRONIC CIRCUIT

Here, we present the complete numerical simulation results for Lorenz63 and the double-scroll electronic circuit. In Fig. D1 and D2, the simulation results of Lorenz63 are given. In the right end of Fig. D2, a significant prediction error emerges, validating the boundary of our current strategy. The simulation results of double-scroll electronic circuit are given in Fig. D3 and D4. Similarly, in the right end of Fig. D4, a significant prediction error emerges, validating the boundary of our current strategy.

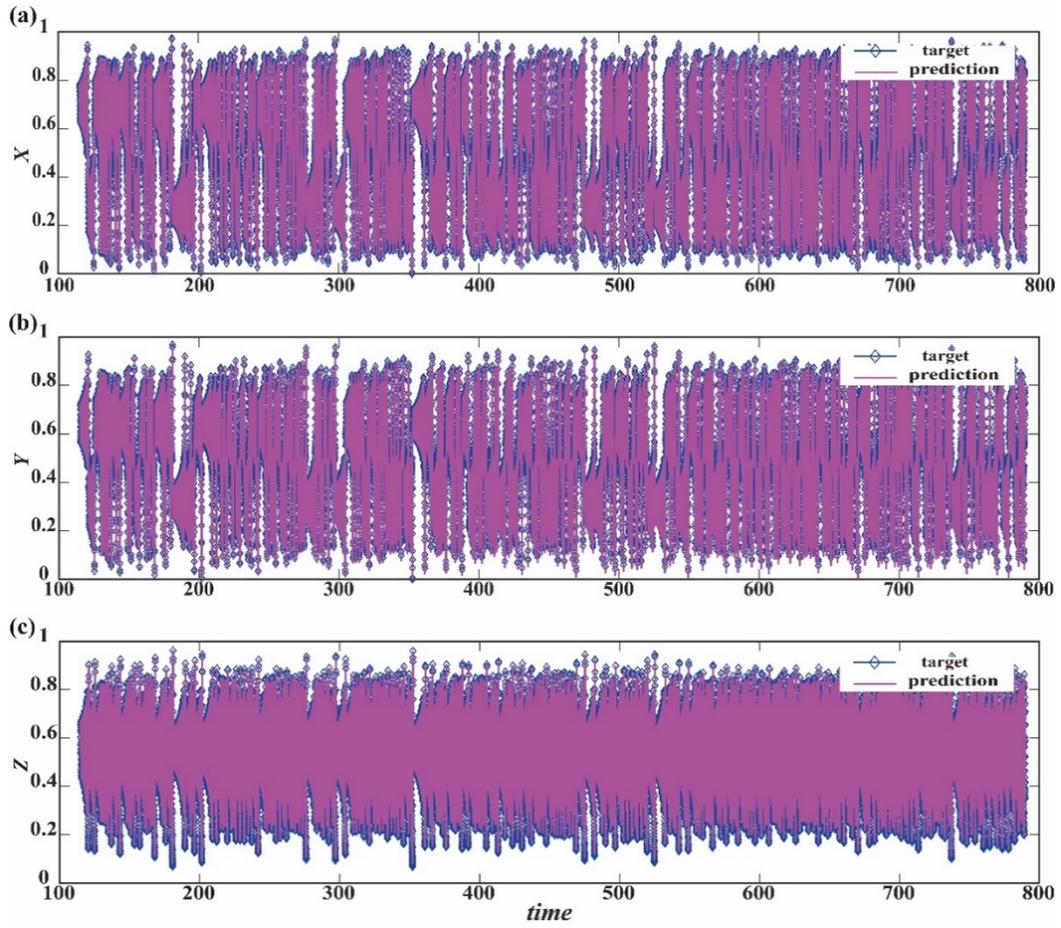

**Fig.D1.** The values of $X, Y, Z$ as a function of time about Lorenz63. (a), (b), (c). The values of $X, Y, Z$ of the target (blue) and predication(purple) data during test phase.

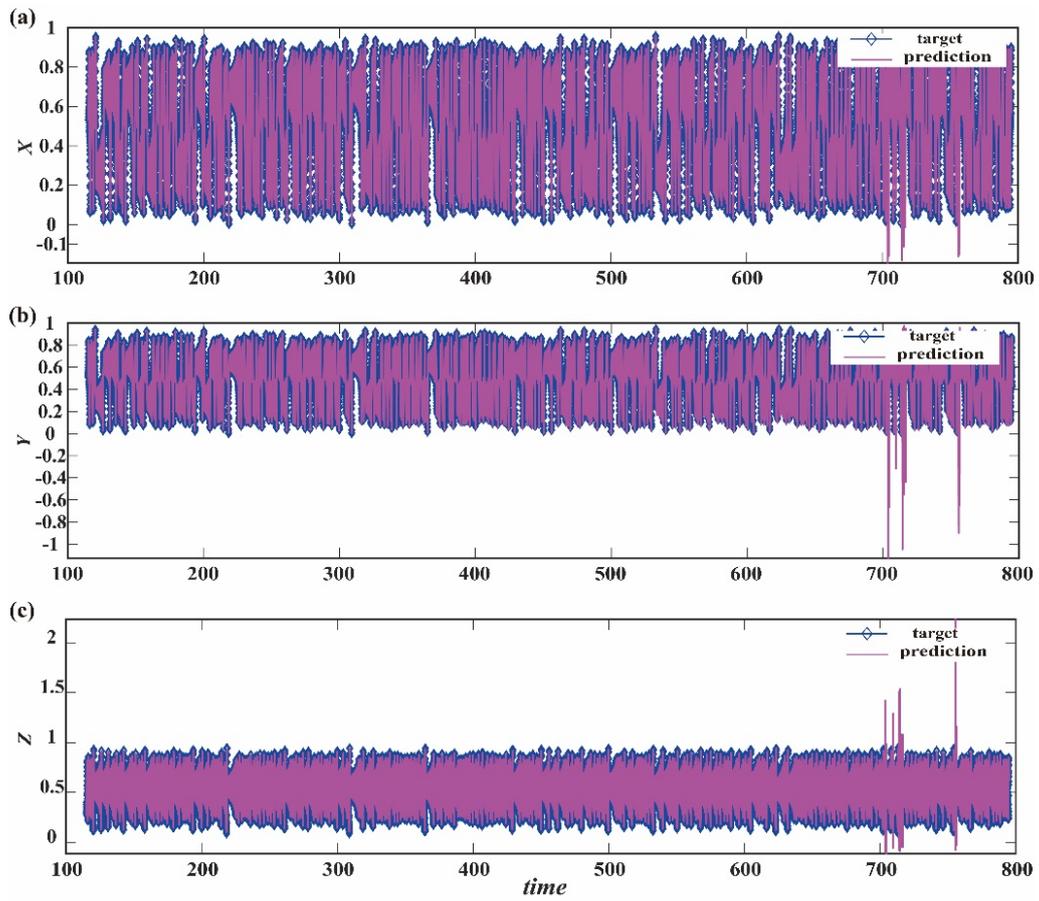

**Fig. D2.** The values of $X, Y, Z$ as a function of time about Lorenz63. (a), (b), (c). The values of $X, Y, Z$ of target (blue) and predication(purple) data during test phase.

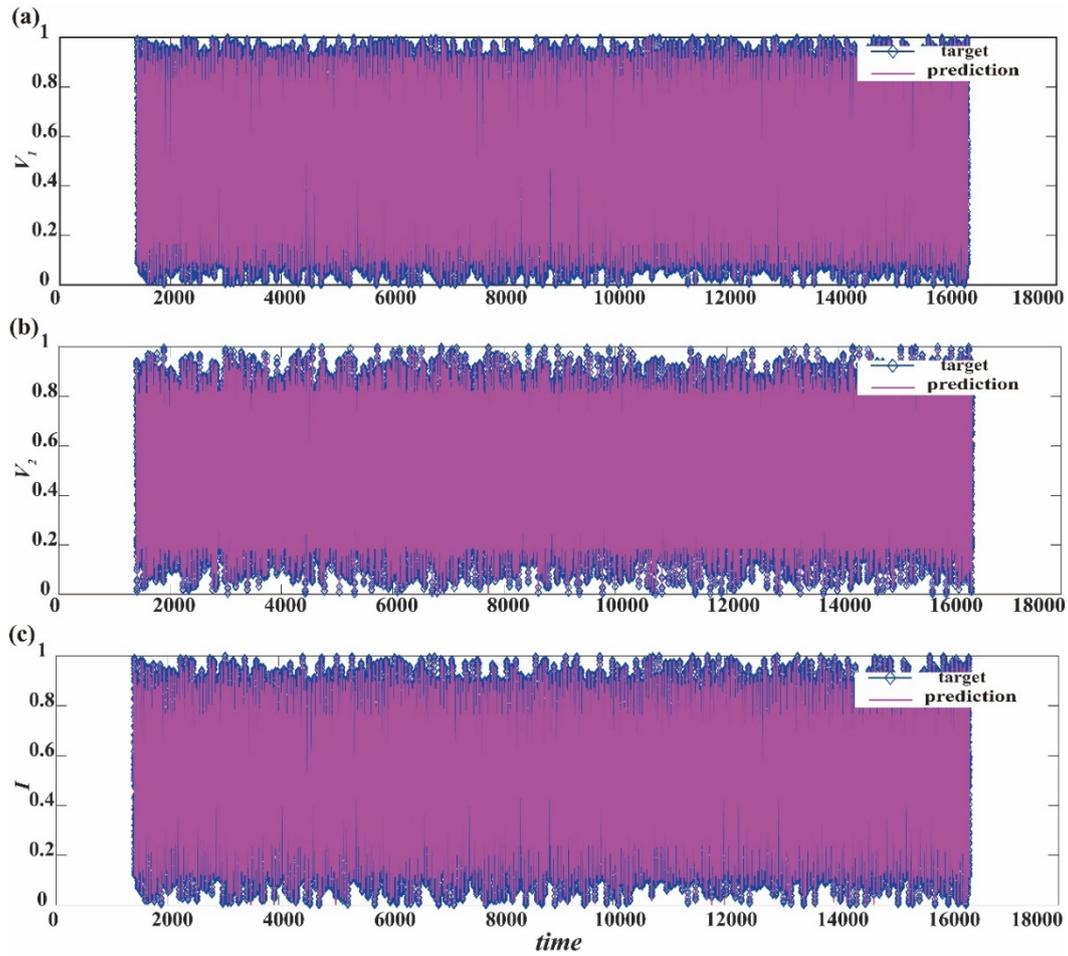

**Fig.D3.** The values of $V_1, V_2, I$ as a function of time about double-scroll electronic circuit. (a), (b),

(c). The values of $V_1, V_2, I$ of the target (blue) and predication(purple) data during test phase.

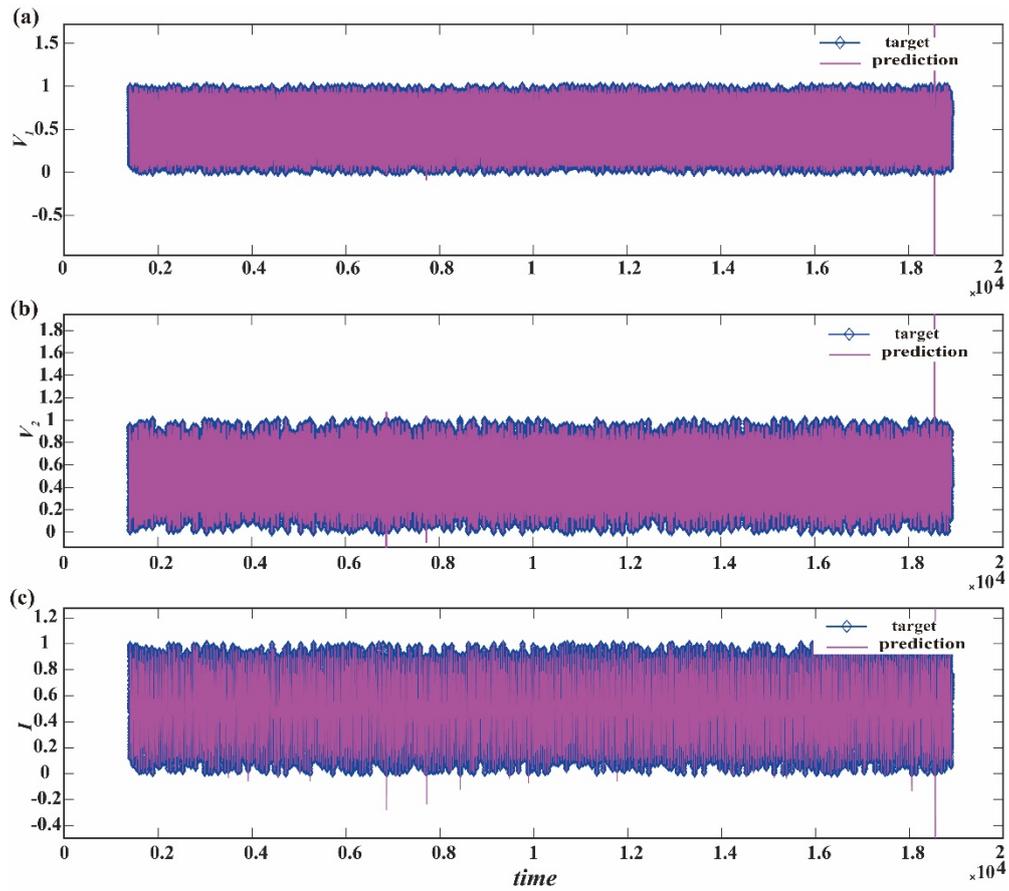

**Fig. D4.** The values of $V_1, V_2, I$ as a function of time about double-scroll electronic circuit. (a), (b),

(c). The values of $V_1, V_2, I$ of target (blue) and predication(purple) data during test phase.

## APPENDIX E: POWER SPECTRUM DENSITY(PSD)

As mentioned in the main text, we provide two types of PSDs to characterize whether the structure of the predicted dynamics matches the target. The first type of PSD is based on the results of Ref. [58], shown in Fig. E1. The second type of PSD is based on the results of Ref. [59], shown in Fig. E2. From both Fig. E1 and Fig. E2, we can see that the target data and prediction data matches well.

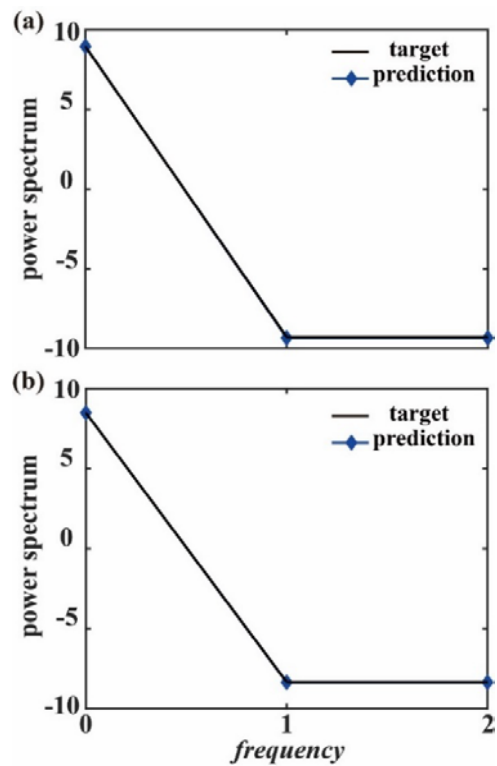

**Fig.E1.** The PSD of the target and the prediction as a function of frequency in (a) the Lorenz63 system and (b) the double-scroll electronic circuit system, according to Ref. [58]. The target results are colored black, and the prediction results are colored blue.

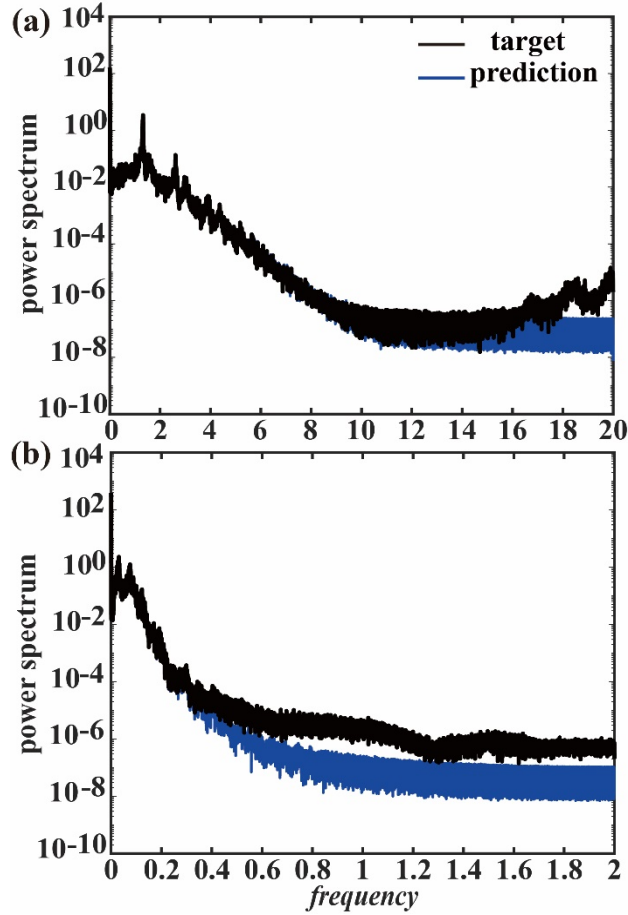

**Fig. E2.** The PSD of the target and the prediction as a function of frequency in (a) the Lorenz63 system and (b) the double-scroll electronic circuit system, according to Ref. [59]. The target results are colored black, and the prediction results are colored blue.

## APPENDIX F: EXPERIMENTAL FEASIBILITY

We have analyzed the experimental feasibility and the hardness of our approach from four aspects: Input encoding, Input map, Creating $H_0$, and Tomography.

a) Input encoding: The preparation of the input state could be not trivial. The results of Ref. [72] indicates that it will generally take $O(2^n)$ steps to prepare an arbitrary $n$-dimensional quantum state, which is the time complexity of a quantum state preparation algorithm. However, the real preparation time is highly dependent on the specific property of physical setup of the system, and can be largely reduced by parallel setups. For example, if the system in Ref. [52] is considered for the implementation, the input state, i.e., the input mode of photons can be tuned parallelly. Although the steps are still generally $O(2^n)$, but the experimental time will be rather short. Of cause, the strategy can be generalized to arbitrary dimensions.

b) Input map: Eq. (2) gives the particular input strategy of our proposal. Here, we explain it more directly by using a 2-qubit quantum state. We assume the initial state of the system is

$$\rho = \begin{bmatrix} \psi_1\psi_1^* & \psi_1\psi_2^* & \psi_1\psi_3^* & \psi_1\psi_4^* \\ \psi_2\psi_1^* & \psi_2\psi_2^* & \psi_2\psi_3^* & \psi_2\psi_4^* \\ \psi_3\psi_1^* & \psi_3\psi_2^* & \psi_3\psi_3^* & \psi_3\psi_4^* \\ \psi_4\psi_1^* & \psi_4\psi_2^* & \psi_4\psi_3^* & \psi_4\psi_4^* \end{bmatrix}. \tag{G1}$$

The above state can be implemented by the waveguide system described in Ref. [52], and four waveguides are required. The light intensities of each waveguide are $\psi_1\psi_1^*$, $\psi_2\psi_2^*$, $\psi_3\psi_3^*$, and $\psi_4\psi_4^*$. The phase of each waveguide mode can be tuned by introducing phase changing materials. As we explained, the coupling coefficients between any two waveguides corresponds to the off-diagonal elements of $H_0$. The process of Eq. (2) under the above setup is as follows. By taking the partial trace of the first qubit, one has

$tr_1(\rho) = \begin{bmatrix} \psi_1\psi_1^* + \psi_3\psi_3^* & \psi_1\psi_2^* + \psi_3\psi_4^* \\ \psi_2\psi_1^* + \psi_4\psi_3^* & \psi_2\psi_2^* + \psi_4\psi_4^* \end{bmatrix}$. A direct implementation is combining the output of the first (second) waveguide and the third (fourth) waveguide into a single propagation mode through interferometers, and adjusting the phase of the mode correspondingly. The intensities of the resulting beams are $\psi_1\psi_1^* + \psi_3\psi_3^*$ and $\psi_2\psi_2^* + \psi_4\psi_4^*$ respectively. An important condition for achieving so is tuning the coherence of the four modes before the combination, setting $\psi_1(\psi_3)$ and $\psi_2(\psi_4)$ to be incoherent. Meanwhile, coherence of $\psi_1(\psi_2)$ and $\psi_3(\psi_4)$ are required to be kept. Next, suppose the input state is $\rho_{in} = \begin{bmatrix} \alpha_1\alpha_1^* & \alpha_1\alpha_2^* \\ \alpha_2\alpha_1^* & \alpha_2\alpha_2^* \end{bmatrix}$. After performing partial trace, if one takes the tensor product of $\rho_{in}$ and $tr_1(\rho)$, one has

$$\rho_{in} \otimes tr_1(\rho) =$$

$$\begin{bmatrix} \alpha_1 \alpha_1^* \left( \psi_1 \psi_1^* + \psi_3 \psi_3^* \right) & \alpha_1 \alpha_1^* \left( \psi_1 \psi_2^* + \psi_3 \psi_4^* \right) & \alpha_1 \alpha_2^* \left( \psi_1 \psi_1^* + \psi_3 \psi_3^* \right) & \alpha_1 \alpha_2^* \left( \psi_1 \psi_2^* + \psi_3 \psi_4^* \right) \\ \alpha_1 \alpha_1^* \left( \psi_2 \psi_1^* + \psi_4 \psi_3^* \right) & \alpha_1 \alpha_1^* \left( \psi_2 \psi_2^* + \psi_4 \psi_4^* \right) & \alpha_1 \alpha_2^* \left( \psi_2 \psi_1^* + \psi_4 \psi_3^* \right) & \alpha_1 \alpha_2^* \left( \psi_2 \psi_2^* + \psi_4 \psi_4^* \right) \\ \alpha_2 \alpha_1^* \left( \psi_1 \psi_1^* + \psi_3 \psi_3^* \right) & \alpha_2 \alpha_1^* \left( \psi_1 \psi_2^* + \psi_3 \psi_4^* \right) & \alpha_2 \alpha_2^* \left( \psi_1 \psi_1^* + \psi_3 \psi_3^* \right) & \alpha_2 \alpha_2^* \left( \psi_1 \psi_2^* + \psi_3 \psi_4^* \right) \\ \alpha_2 \alpha_1^* \left( \psi_2 \psi_1^* + \psi_4 \psi_3^* \right) & \alpha_2 \alpha_1^* \left( \psi_2 \psi_2^* + \psi_4 \psi_4^* \right) & \alpha_2 \alpha_2^* \left( \psi_2 \psi_1^* + \psi_4 \psi_3^* \right) & \alpha_2 \alpha_2^* \left( \psi_2 \psi_2^* + \psi_4 \psi_4^* \right) \end{bmatrix}.$$

In the waveguide system based on Ref. [52], this corresponds to splitting the two propagation modes obtained above to generate four modes. Two of them have the intensity $\psi_1 \psi_1^* + \psi_3 \psi_3^*$, and the other two have the intensity $\psi_2 \psi_2^* + \psi_4 \psi_4^*$. Then, modulate the intensities of the four beams to $\alpha_1 \alpha_1^* \left( \psi_1 \psi_1^* + \psi_3 \psi_3^* \right)$, $\alpha_1 \alpha_1^* \left( \psi_2 \psi_2^* + \psi_4 \psi_4^* \right)$, $\alpha_2 \alpha_2^* \left( \psi_1 \psi_1^* + \psi_3 \psi_3^* \right)$, and $\alpha_2 \alpha_2^* \left( \psi_2 \psi_2^* + \psi_4 \psi_4^* \right)$. Notice that the components of $\alpha_1$ and $\alpha_2$ are required to be coherent. This can be achieved by using the statistical property of light field. Additionally, the phase of the modes at this stage can also be adjusted by introducing phase changing materials.

c) Creating $H_0$: The time of creating $H_0$ by experimental platform depends on the specific setups. If the tuning of $H_0$ is implemented by Mach-Zehnder interferometer as in the reply to the first comment. The experimental time involves the modulation and stabilization of interferometers. In fact, it relies on the particular strategy for phase modulation (via thermal tuning or others), and can also be performed parallelly.

d) Tomography: We employ quantum tomography rather partial measurements (such as Pauli-$Z$ measure in Ref. [34]) in each evolution step to gather the output. Although such a process has a high complexity, it shall not be involved in our consideration. This is because the main purpose of our proposal is to gather data for predicting the chaotic dynamics without direct solving the original equation. Chaotic dynamics is obviously complicated, and predicting the behavior naturally necessities generating complicated patterns. The key point of our proposal is that one can merely measure a system to obtain a relatively long prediction of such a complicated dynamic. Hence, our proposal generates data of complicated patterns, and extracting the generated data also requires large number of measurements. Therefore, the complexity introduced by topographical measurements is a natural consequence in the above prediction problem. Actually, our proposal uses a recurrent setup to achieve a low prediction error and longer prediction time, which is an advance along the line of research.

## APPENDIX G: THE LARGEST LYAPUNOV EXPONENT

First, to measure the largest Lyapunov exponent of the data generated by Lorenz63, we uniformly sample the data. Then, the time delay parameter for effectively reconstructing a 3-dimensional phase space of the sampled data is estimated using average mutual information (AMI), defined by $\sum_{i=1}^{L} p(x_i, x_{i+T}) \log_2[p(x_i, x_{i+T})/p(x_i)p(x_{i+T})]$. $L$ represents the total number of samples, $p(x_i)$ and $p(x_i, x_{i+T})$ are given by the distribution of samples. Hence, the time delay parameter is given by $T$ when the AMI reaches its first local minimum.

Next, using the time delay parameter obtained in the previous step, we create the average logarithmic divergence versus expansion step plot for the data. At the same time, set a sufficiently large expansion range to capture all expansion steps.

Finally, the Lyapunov exponent for the entire expansion range is calculated by Rosenstein algorithm Ref. [73]. By linearly fitting the line to the best with the original data, a new expansion range can be obtained. The new values of the expansion ranges are utilized to find the largest Lyapunov exponent for the Lorenz63 system Ref. [74].

(2020).